\documentclass[twoside]{sfm}

\input{sf.def}

\begin{document}

\def\llm{{\sc LLmodels}}
\def\atl{{\sc ATLAS9}}
\def\aatl{{\sc ATLAS12}}
\def\starsp{{\sc STARSP}}
\def\aur{$\Theta$~Aur}
\def\logg{\log g}
\def\teff{T_{\rm eff}}
\def\tauros{\tau_{\rm Ross}}
\def\kms{km\,s$^{-1}$}
\def\bz{$\langle B_{\rm z} \rangle$}
\def\degr{^\circ}
\def\aaps{A\&AS}
\def\aap{A\&A}
\def\apjs{ApJS}
\def\apj{ApJ}
\def\rmxaa{Rev. Mexicana Astron. Astrofis.}
\def\mnras{MNRAS}
\def\actaa{Acta Astron.}
\newcommand{\Tef}{T$_{\rm eff}$~}
\newcommand{\Vt}{$V_t$}
\newcommand{\CC}{$^{12}$C/$^{13}$C~}
\newcommand{\CDC}{$^{12}$C/$^{13}$C~}
\newcommand{\vsini}{\textit{V}sin\textit{i}}
\newcommand{\hbeta}{H$_\beta$}

\pagebreak

\thispagestyle{titlehead}

\setcounter{section}{0}
\setcounter{figure}{0}
\setcounter{table}{0}

\markboth{K{\i}l{\i}\c{c}o\u{g}lu et al.}{The open cluster M6}

\titl{Elemental Abundance Analysis of the Early Type Members of the Open Cluster M6: Preliminary Results}{K{\i}l{\i}\c{c}o\u{g}lu T.$^1$, Monier R.$^{2,3}$, Fossati L.$^4$}
{$^1$Department of Astronomy and Space Sciences, Faculty of Science, Ankara University, Turkey, email: {\tt tkilicoglu@ankara.edu.tr} \\
 $^2$Lagrange Laboratory, Nice University, France\\
 $^3$LESIA, Paris Observatory, Meudon, France\\
 $^4$Argelander-Institute for Astronomy, Bonn University, Germany}

\abstre{
Differences in chemical composition among main sequence stars within a given cluster are probably due to differences in their masses and other effects such as radiative diffusion, magnetic field, rotation, mixing mechanisms, mass loss, accretion and multiplicity. The early type main-sequence members of open clusters of different ages allow to study the competition between radiative diffusion and mixing mechanisms. We have analysed low and high resolution spectra covering the spectral range 4500 - 5840 \AA\ of late B, A, and F type members of the open Cluster M6 (age $\sim$100 Myr). The spectra were obtained using the FLAMES/GIRAFFE spectrograph mounted at UT2, the 8 meter class VLT telescope. The effective temperatures, surface gravities and microturbulent velocities of the stars were derived  using both photometric and spectral methods. We have also performed a chemical abundance analysis using synthetic spectra. The abundances of the elements were determined for C, O, Mg, Si, Ca, Sc, Ti, Cr, Mn, Fe, Ni, Y, Ba. The star-to-star variations in elemental abundances among the members of the open cluster M6 were discussed.}

\baselineskip 12pt

\section{Introduction}

The elemental abundance patterns of F-type main sequence stars, having no strong magnetic field, usually reflect the chemical compositions of their birthplaces as their interiors are mixed by convection. In contrast, the A-type main sequence stars may strongly depart from this initial composition and display significant star-to-star differences in elemental abundance due to the occurrence of diffusion and/or various kind of mixing mechanisms in their radiative envelopes. Open clusters are useful laboratories to set constraints on predictions of radiative diffusion and other kind of mixing mechanisms (\cite{trm}, \cite{rmt}). Elemental abundance analyses of both normal and chemically peculiar stars in several open cluster have been performed by several authors, e.g. \cite{vm}, \cite{m}, \cite{gm}, \cite{gmr}, \cite{vcs}, \cite{gvmf}, \cite{ffb}. In order to understand the nature of these mixing mechanisms, elemental abundance analyses of A and F stars in open clusters of various ages are highly desired. 

M6 is a bright (mean integrated \textit{V}=4$^m$.2) and young (age $\sim$100 Myr) southern open cluster observable. The UBV Johnson photometric observations of the cluster can be retrieved from \cite{r}, \cite{e}, \cite{t}, \cite{a}, and \cite{v}. The likely members of the cluster were first numbered by \cite{r}, and then extended by \cite{a}. The photometric metallicity of [Fe/H] = 0.07 dex was derived from photometry by \cite{c}. Str\"omgren and Geneva 7color Photometry of the cluster were performed by \cite{s} and \cite{n}, respectively. Chemically peculiar stars were searched for in the cluster by \cite{ms} and \cite{p} using $\Delta$a photoelectric/CCD measurements. Except for the photographic plates obtained by \cite{r}, there is no spectroscopic data available for cluster members. A few magnetic CP stars were studied spectroscopically, e.g. \cite{blb}.

The incentive of this paper is to report on the abundance determinations of 13 elements in 9 A and 8 F dwarfs in the open cluster M6, and search for correlations between their elemental abundances and stellar parameters: \Tef , $\vsini$.

\section{Observations}\label{giraffe}

Spectra of 17 stars in the region of the open cluster M6 were obtained by the fiber-fed FLAMES/GIRAFFE spectrograph mounted on the 8.2 meter VLT telescope. A low resolution (R$\sim$7500) region (L479.7) was selected to acquire \hbeta\ profile and many prominent metallic lines of Fe, Cr, Ti and, Mn. Two high resolution (R$\sim$25000) regions (H525.8B and H572.8) including many metallic lines were also selected to derive more elemental abundances with better accuracy. The fundamental parameters of the B, A and, F-type probable members rotating less than 150 \kms\ are collected in Table~\ref{fundamental}.

\begin{table}[!t]
\centering
\caption{Derived fundamental parameters of the members} \label{fundamental} \tabcolsep1.2mm
\begin{tabular}{cccccccc}
\hline\noalign{\smallskip}
 & & \multicolumn{2}{c}{Photometry} & \multicolumn{2}{c}{Final} & & \\
 \multicolumn{1}{c}{No.} & Other ident. & \Tef &  \multicolumn{1}{c}{$\logg$} & \Tef &\multicolumn{1}{c}{$\logg$} &
\vsini & Sp. T. \\
 & & \multicolumn{1}{c}{(K)} & \multicolumn{1}{c}{(dex)} & \multicolumn{1}{c}{(K)} & \multicolumn{1}{c}{(dex)} & \multicolumn{1}{c}{(\kms)}\\
\hline
\multicolumn{1}{c}{1} & \multicolumn{1}{c}{2} & 3 & \multicolumn{1}{c}{4} & \multicolumn{1}{c}{5} & \multicolumn{1}{c}{6}
 & \multicolumn{1}{c}{7} & \multicolumn{1}{c}{8} \\
\hline \noalign{\smallskip}
20 & HD 318101 & 15864 & 4.04 & 15400 & 4.00 & 31.0 & B5 \\ 
115 & HD 160167 & 13032 & 4.19 & 12200 & 4.00 & 120.0 & B8 \\ 
99 & HD 318126 & 12369 & 4.33 & 12000 & 4.10 & 55.0 & B8 \\ 
17 & HD 160392 & 12090 & 4.19 & 11500 & 4.10 & 48.0 & B8 \\ 
- & HD 318112 & 11521* & - & 11100 & 4.20 & 150.0 & B9 \\ 
29 & HD 318099 & 11199 & 4.21 & 10900 & 4.30 & 140.0 & B9 \\ 
25 & HD 318093 & 10368 & 4.13 & 10400 & 4.30 & 75.0 & B9 \\ 
63 & HD 318118 & 10116 & 4.27 & 9700 & 4.20 & 100.0 & A0 \\ 
47 & CD-32 13109  & 9396 & 4.30 & 9400 & 4.20 & 5.1 & A1 \\ 
71 & CD-32 13106 & 9418 & 4.14 & 8900 & 4.00 & 135.0 & A2 \\ 
5 & HD 318091 & 9079 & 4.21 & 8700 & 4.00 & 62.0 & A3 \\ 
- & CPD-32 4724 & - & - & 8500 & 4.20 & 70.0 & A4 \\ 
48 & CPD-32 4713 & 8387 & 4.07 & 8250 & 4.05 & 120.0 & A5 \\ 
53 & CD-32 13089 & 7991* & - & 8150 & 3.60 & 75.0 & A5 \\ 
130 & CD-32 13148 & 8079 & 3.94 & 8000 & 4.10 & 120.0 & A6 \\ 
- & HD 318103 & 7470* & - & 7800 & 4.20 & 115.0 & A7 \\ 
14 & GSC 7380-0766 & 7230 & 4.76 & 7200 & 4.20 & 80.0 & F1 \\ 
118 & CPD-32 4693 & 7044 & 4.60 & 7150 & 4.10 & 64.0 & F1 \\ 
66 & GSC 7380-0986 & 6532 & 4.45 & 6850 & 4.40 & 17.9 & F2 \\ 
69 & GSC 7380-1211 & 6382 & 3.81 & 6850 & 3.80 & 57.0 & F3 \\ 
18 & GSC 7380-0206 & 6045 & 3.22 & 6700 & 2.70 & 110.0 & F3 \\ 
6 & GSC 7380-1170 & 6312* & - & 6700 & 4.50 & 12.9 & F3 \\ 
27 & GSC 7380-0339 & 6147* & - & 6650 & 4.80 & 7.0 & F5 \\ 
E32c & GSC 7380-1363 & 6279* & - & 6500 & 4.20 & 22.0 & F6 \\ 

\noalign{\smallskip}\hline
\multicolumn{8}{l}{Notes.} \\
\multicolumn{8}{l}{$^1$ star no. from \cite{r} and \cite{a}, $^2$ other identifier} \\
\multicolumn{8}{l}{$^{3,4}$ atmospheric parameters derived from Geneva 7color photometry,} \\
\multicolumn{8}{l}{$^{5,6}$ atmospheric parameters derived by spectroscopy,} \\
\multicolumn{8}{l}{$^7$ projected rotational velocity, $^8$ spectral type estimated from \Tef} \\
\multicolumn{8}{l}{* effective temperatures estimated from Johnson (B-V)$_0$ color} \\
\end{tabular}
\end{table}

\section{Abundance analysis}\label{analysis}

In order to derive the abundances of various elements, we used synthetic spectrum method. Indeed most members of the cluster are fast rotating stars and the classical equivalent width measurement technique is not suitable to determine the abundances. For each star, a model atmosphere was computed using ATLAS9 (\cite{k},\cite{sbc},\cite{sb}) with the prescriptions of mixing-length ratios provided by \cite{sm}. The synthetic spectra were computed using SYNSPEC48 (\cite{hl}). The effective temperatures and the logarithm of surface gravities of the stars were estimated first by Johnson and Geneva 7color photometric systems (Table~\ref{fundamental}) with the calibrations of \cite{gr} and \cite{knk}, respectively. We did not use Str\"omgren Photometry since M6 has not been observed in Str\"omgren \hbeta\ filter. We then compared the observed \hbeta\ lines with the predicted \hbeta\ profiles to improve the atmospheric parameters. We have also used ionization and excitation equilibria to check the atmospheric parameters for few slow rotating members. The error on \Tef is around $\pm$100 K for F-type stars, and increases up to 300 K for hot stars. The error on $\logg$ is $\pm$0.2 dex in most cases, and it decreases to $\pm$0.1 dex for hottest stars. The rotation velocities were derived by adjusting synthetic line profiles to the unblended weak metallic lines (mostly Fe II, Cr II, and Ti II) of stars. The atomic list of lines was first constructed from gfhyperall.dat provided by \cite{k}, and then updated by using VALD (\cite{vald}),NIST and recent publications. Hyperfine structure was taken into account. The abundances of the elements were derived by iteratively adjusting the synthetic spectra to the normalized spectra, and minimizing the chi-square of the models to the observations. We did not use any automated iteration to avoid of erroneous abundance determinations caused by highly blended lines in fast rotators. 

\begin{figure}[!t]
\begin{center}
\hbox{
 \includegraphics[clip=true]{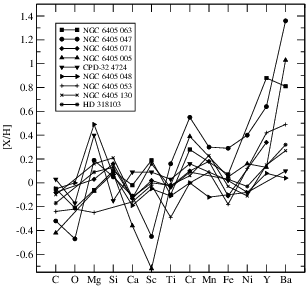} \ 
 \includegraphics[clip=true]{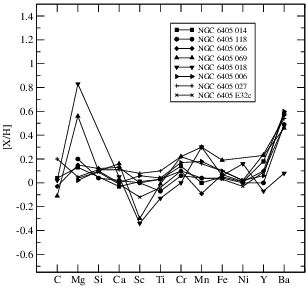}

}
\vspace{-2mm}
\caption[]{Abundances of the elements with respect to the Sun for A-type (left panel) and F-type (right panel) members. }
\label{abundances}
\end{center}
\end{figure}

\section{Conclusions}

The abundances of 13 elements for A-type stars, and 12 elements for F-type stars were derived. For B-type stars, only few lines of a limited number of the elements are observed in this wavelength region. Additional spectra in the blue would help specify more abundances for these hot stars. Star-to-star abundance variations are found for A-type stars and they are larger than for F-type stars as depicted in Figure\,\ref{abundances}. This scatter in abundances is especially large for Mg, Cr, Mn, Fe, Ni, Y, and Ba, which was already reported by \cite{gm}, \cite{gvmf}. The origin of this scatter is not clear yet but probably reflects the competition between radiative separation and mixing by different rotation rates. The steadier chemical pattern of the F-type stars provides the mean abundance of the cluster which must be close to the original abundances. We could not find any correlation between [Fe/H] and \vsini, nor \Tef . Similar correlations was searched for by \cite{ffb} as well. A comparison between these derived abundances, the surface abundances predicted by evolutionary models treating self-consistently radiative and turbulent diffusion for the age of M6, and a likely initial chemical mix will help us to understand the competition between radiative diffusion and mixing mechanisms in this young open cluster.

\begin{figure}[!t]
\begin{center}
\hbox{
 \includegraphics[clip=true]{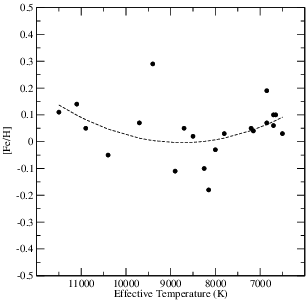} \ 
 \includegraphics[clip=true]{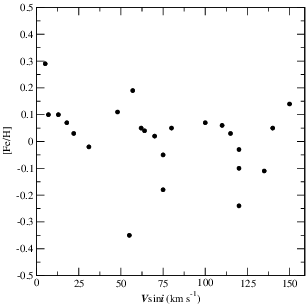}

}
\vspace{-2mm}
\caption[]{Iron abundance versus \Tef (left panel) and \vsini\ (right panel) for program stars}
\label{relation}
\end{center}
\end{figure}

\bigskip
{\it Acknowledgements.} We kindly thank Pierre North for making his code CALIB available. This research was supported by the Scientific and Technological Research Council of Turkey (T\"UB\.{I}TAK, 1001-112T119). We made use of the SIMBAD, WEBDA, VALD, and NIST databases.

\end{document}